\documentclass[12pt]{article}
\usepackage{epsfig}
\usepackage{pst-plot,url,epsf}
\usepackage{t1enc}
\usepackage{placeins}
\newcommand{\beq}{\begin{equation}}
\newcommand{\eeq}{\end{equation}}
\newcommand{\bea}{\begin{eqnarray}}
\newcommand{\eea}{\end{eqnarray}}

\newcommand{\epm}{e^+e^-}
\newcommand{\nn}{\nonumber}

\newcommand{\ra}{\rightarrow}
\def\earr{\end{array}}
\def\barr#1{\begin{array}{#1}}

\begin{document}
\thispagestyle{empty}
\begin{flushright}
July 2015\\
Revised version:\\
September 2015\\
\vspace*{1.5cm}
\end{flushright}
\begin{center}
{\LARGE\bf Probing the top--Higgs coupling through the secondary lepton 
distributions in the associated production of the top-quark pair and Higgs 
boson at the LHC}\\
\vspace*{2cm}
Karol Ko\l odziej\footnote{E-mail: karol.kolodziej@us.edu.pl} and
Aleksandra S\l apik\footnote{E-mail: aleksandra.slapik@gmail.com}
\\[1cm]
{\small\it
Institute of Physics, University of Silesia\\ 
ul. Uniwersytecka 4, PL-40007 Katowice, Poland}\\
\vspace*{4.5cm}
{\bf Abstract}\\
\end{center}
We complement the analysis of the anomalous top--Higgs coupling effects on 
the secondary lepton distributions in the associated production of the 
top-quark pair and Higgs boson in proton--proton collisions at the LHC 
of the former work by one of the present authors by taking into account the 
quark--antiquark production mechanism. We also present 
simple arguments which explain why the effects of the scalar and pseudoscalar
anomalous couplings on the unpolarized cross section of the process are 
completely insensitive to the sign of either of them.

\vfill

\newpage

\section{Introduction}

Determination of the coupling of the recently discovered Higgs boson 
\cite{hdiscovery} to the top quark currently belongs to one of the most 
challenging tasks of the high energy experimental physics. Measurement of 
the associated production of the top quark pair and Higgs boson in the clean 
experimental environment of $\epm$ collisions was considered in this context 
already more than two decades ago \cite{gtth1}, \cite{gtth2}, but different 
projects of the high energy $\epm$ collider \cite{ILC}--\cite{CEPC2},
despite some of them being more or less intensively discussed for years, are 
still at a rather early stage of TDR. However, if the LHC performance in next
runs is as excellent as it was in run 1 we may expect that the 
process
\bea
\label{pptth}
pp \;\ra\; t \bar t H 
\eea
the search for which, based on run 1 data, were already reported by both the 
CMS \cite{tthCMS} and ATLAS \cite{tthATLAS} collaborations, will be measured 
quite precisely. This is why in the past few years the associated 
production of the top quark pair and Higgs boson has 
invoked quite some interest also from a theoretical side, see, 
e.g., \cite{kkss}--\cite{buckley}.

\begin{figure}[htb]
\centerline{
\epsfig{file=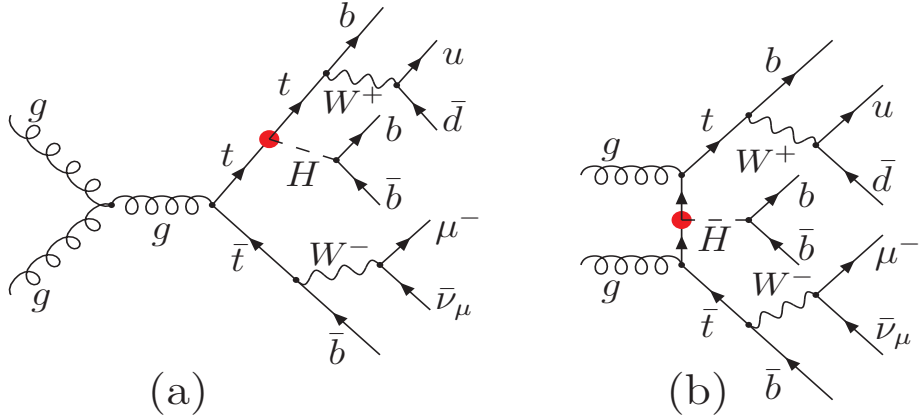,  width=120mm, height=55mm}}
\caption{Feynman diagrams of $t \bar t H$ production in process
(\ref{ggbbbudbmn}). Blobs indicate the Higgs--top coupling.}
\label{ggsig}
\end{figure}

It was shown in Ref.~\cite{kkjhep} that the distributions in rapidity and angles 
of the secondary lepton that can be produced in the decay of $\bar t$-quark of 
process (\ref{pptth}) are quite sensitive to modifications of the top--Higgs 
coupling. Actually, only the gluon fusion mechanism of $t \bar t H$ 
production, which is dominant at the LHC energies, and one specific decay 
channel:
$t\to bW^+\to b u\bar d$, $\bar t\to \bar bW^-\to \bar b\mu^-\bar \nu_{\mu}$ and
$h\to b\bar b$, were taken into account in Ref.~\cite{kkjhep}, i.e., the
following hard parton scattering processes
\bea
\label{ggbbbudbmn}
gg \;\ra\; b u \bar{d}\, \bar b \mu^- \bar \nu_{\mu} b \bar b,
\eea
was considered. There are $67\,300$ Feynman diagrams of process 
(\ref{ggbbbudbmn}) already 
in the leading order (LO) of the standard model (SM) in the unitary gauge, if the
Cabibbo-Kobayashi-Maskawa mixing and masses 
smaller than the $b$-quark mass $m_b$ are neglected.
At the same time there are only 32 Feynman diagrams which contribute to the 
signal cross section of $t \bar t H$ production, two of which are shown in 
Fig.~\ref{ggsig}. The remaining 30 signal diagrams are obtained from those 
depicted by attaching the Higgs boson line of $Hb\bar b$-vertex to the other $t$- 
or $\bar t$-quark line, or interchanging the $b$ and $\bar b$ quarks 
in Figs.~\ref{ggsig}(a) and \ref{ggsig}(b) and interchanging the two gluons in 
Fig.~\ref{ggsig}(b). The diagrams with the Higgs boson line of  $Hb\bar b$-vertex 
attached to either the $b$- or $\bar b$-quark line are not counted here, 
as their contribution to the $t \bar t H$ production signal is suppressed by the 
mass ratio $m_b/m_t$. The effects caused by modifications of the scalar 
and pseudoscalar couplings of the Higgs boson to top quark were clearly 
visible in the $t \bar t H$ production signal cross section, but they were to 
large degree obscured by the interference of the $t \bar t H$ production signal 
diagrams with the diagrams of irreducible off resonance background.

\begin{figure}[htb]
\centerline{
\epsfig{file=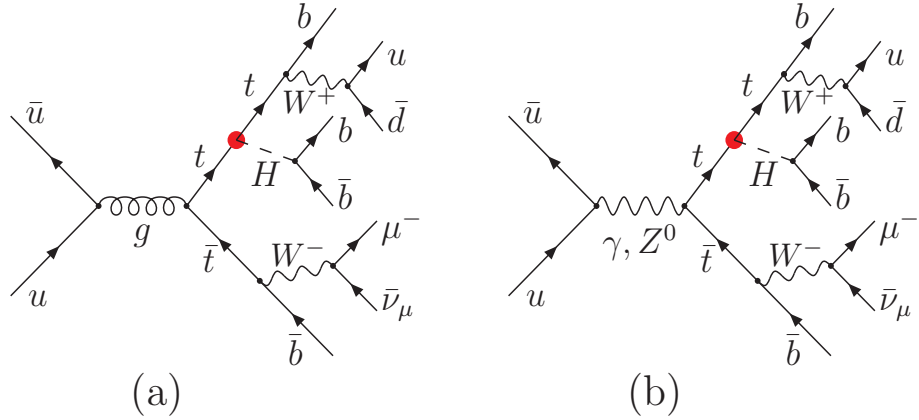,  width=120mm, height=55mm}}
\caption{Feynman diagrams of $t \bar t H$ production in the process
$u \bar{u} \;\to\; b u \bar{d} \bar{b} \mu^{-} \bar{\nu}_{\mu} b \bar{b}$. 
Blobs indicate the Higgs--top coupling.}
\label{uusig}
\end{figure}

In the present work, we complement the analysis of the influence of the 
anomalous Higgs boson coupling to top quark on the secondary lepton 
distributions in the process of associated production of the top quark pair 
and Higgs boson in proton--proton collisions at the LHC of Ref.~\cite{kkjhep}
by taking into account the quark--antiquark annihilation hard scattering 
processes with the same final state as that of process (\ref{ggbbbudbmn}):
\bea 
q \bar{q} \;\to\; b u \bar{d}\, \bar{b} \mu^{-} \bar{\nu}_{\mu} b \bar{b},
\label{qqbar}
\eea
with $q=u,d$. To be more specific, we take into account $u\bar u$-,  
$\bar u u$-, $d\bar d$- and $\bar d d$-scattering processes. 
Under the same assumptions as those made above for process 
(\ref{ggbbbudbmn}), there are $78\,068$ Feynman diagrams in the LO of SM for 
each of the $q\bar q$-scattering processes considered. However, only $24$ of 
them contribute to the signal of the $t \bar{t} H$ production. Examples of the 
signal diagrams of the process of $u\bar u$-scattering to the final state
of process (\ref{qqbar}) are shown in Fig.~\ref{uusig}. 
The other signal diagrams can be obtained by attaching the Higgs 
boson line of the $Hb \bar{b}$-vertex to the other $t$- or $\bar{t}$-quark 
line or interchanging the $b$- and $\bar{b}$-quark lines in the diagrams of 
Fig.~\ref{uusig}. Let us note that another $24$ diagrams which contain the 
Feynman propagators of the $t$-, $\bar{t}$-quark and the Higgs boson at a 
time can be obtained from the signal diagrams just described by the exchange 
of the $u$-quark lines between the initial and final state.
However, they are not treated as the signal diagrams here, because they 
contain the gluon, $Z^0$ or photon propagator in the $t$- or $u$-channel and 
their contribution to the signal cross section is negligible 
anyway, which has been checked by direct computation.

The rest of the article is organized in the following way. The possible effect of 
the anomalous top--Higgs coupling on the unpolarized cross section of the process 
of $t\bar tH$ production at the LHC are analyzed in Section~2, our results
are presented in Section~3 and, finally, some concluding remarks are contained 
in Section~4.

\section{Effects of the anomalous top--Higgs coupling}

The most general top--Higgs coupling is given by the following Lagrangian 
\cite{aguilar}:
\begin{equation}
 \label{Ltth}
\mathcal{L}_{t\bar{t} H} = -g_{t\bar{t}H} \bar{t} (f + i f' \gamma_5) t h,
\end{equation}
where $g_{t\bar tH}=m_t/v$, with $v=(\sqrt{2}G_F)^{-1/2}\simeq 246$~GeV, 
is the top--Higgs Yukawa coupling and the real couplings $f$ and $f'$ describe,
respectively, the scalar and pseudoscalar departures from the purely scalar 
top--Higgs Yukawa coupling of SM, which is reproduced for $f=1$ 
and $f'=0$. The allowed regions of the $(f,f')$ plane, according to the analysis 
of Ref.~\cite{ellis1} performed at the 68 and 95\% confidence level, are plotted 
in Fig.~1 of Ref.~\cite{ellis2}. They are derived from the constraints on
the $Hgg$ and $H\gamma\gamma$ couplings from the Higgs boson 
production and its decay into $\gamma\gamma$, which among others
involve assumptions on the Higgs boson couplings to other fermions and bosons,
and hence are model dependent. Therefore, we will not stick to them in the
next section, where we will illustrate the effects of
$f'$ on the process of associated production of the top quark pair and Higgs 
boson from which the direct constraints on $f$ and $f'$ can be derived.

Let us try to predict the possible effect of the  top--Higgs coupling given by 
Eq.~(\ref{Ltth}) on the unpolarized cross section of the process 
$u \bar{u} \;\to\; b u \bar{d} \bar{b} \mu^{-} \bar{\nu}_{\mu} b \bar{b}$.
To this end, let us consider the amplitudes of two dominant diagrams of the 
$t\bar t H$ production: 
$M_{a}^{(1)}$ of the diagram depicted in Fig.~\ref{uusig}(a) and
$M_{a}^{(2)}$ of the diagram obtained from that of Fig.~\ref{uusig}(a) by attaching
the Higgs boson line to the $\bar t$-quark. They have the following form:
\bea
\label{amp1}
M_{a}^{(1)}&=&g_{t\bar{t}H}h\,\bar u(f+if'\gamma_5)\;
\frac{p_t\!\!\!\!/+q\!\!\!/+M}{(p_t+q)^2-M^2}\;g_s\varepsilon\!\!\!/ \,v,\\
\label{amp2}
M_{a}^{(2)}&=&g_{t\bar{t}H}h\,\bar u g_s\varepsilon\!\!\!/
\;\frac{-p_{\bar t}\!\!\!\!/-q\!\!\!/+M}{(p_{\bar t}+q)^2-M^2}\;(f+if'\gamma_5)\,v,
\eea
where $h$ is a scalar representing a product of the Higgs boson propagator 
carrying the four momentum $q$ with the $Hb\bar b$-vertex, $u$ ($v$) is the Dirac 
spinor representing the off-shell $t$-quark ($\bar t$-quark) of the four 
momentum $p_t$ ($p_{\bar t}$) that decays into the $b$-quark ($\bar b$-quark) 
and off-shell $W^+$ ($W^-$)-boson, $\varepsilon$ is a polarization four vector 
representing the gluon propagator contracted with the $u\bar u g$-vertex, 
$g_s$ is the strong coupling constant and
$M=\sqrt{m_t^2-im_t\Gamma_t}\approx m_t-\frac{i}{2}\Gamma_t$ is a complex 
mass parameter that replaces the mass $m_t$ in the top quark propagator
in order to regularize the pole arising if its denominator approaches zero.
After some simple algebra Eqs.~(\ref{amp1}) and (\ref{amp2}) can be written in the
following form:
\bea
\label{amp21}
M_{a}^{(1)}&=&\frac{g_{t\bar{t}H}g_sh}{(p_t+q)^2-M^2}
\left[f\bar u(p_t\!\!\!\!/+M)\varepsilon\!\!\!/ v
+if'\bar u(-p_t\!\!\!\!/+M)\gamma_5\,\varepsilon\!\!\!/ v
+\bar u(f+if'\gamma_5)q\!\!\!/\varepsilon\!\!\!/ v\right],\\
\label{amp22}
M_{a}^{(2)}&=&\frac{g_{t\bar{t}H}g_sh}{(p_{\bar t}+q)^2-M^2}
\left[f\bar u\,\varepsilon\!\!\!/(-p_{\bar t}\!\!\!\!/+M)v
+if'\bar u\varepsilon\!\!\!/\gamma_5(p_{\bar t}\!\!\!\!/+M) v
-\bar u\varepsilon\!\!\!/q\!\!\!/(f+if'\gamma_5) v\right].
\eea
Now, let us note that, as in the process of $t\bar t H$ production in $\epm$ 
collisions that was considered in Ref.~\cite{kkss}, the dominant 
contribution to the cross section comes from 
the phase space region, where both the $t$-quark and $\bar t$-quark are close to 
their mass shells and hence the off-shell spinors $u$ and $v$ should satisfy 
the following approximate equations:
\bea
\label{Diracu}
\bar u(p_t\!\!\!\!/-m_t)\approx 0, &&\qquad \bar u(p_t\!\!\!\!/+m_t)\approx 2m_t, \\
\label{Diracv}
(p_{\bar t}\!\!\!\!/+m_t)v\approx 0, &&\qquad (p_{\bar t}\!\!\!\!/-m_t)v\approx -2m_t.
\eea
Using Eqs.~(\ref{Diracu}) in (\ref{amp21}) and (\ref{Diracv}) in (\ref{amp22}),
and neglecting terms $\sim\Gamma_t$ in the numerators, we get the following 
approximate expressions for the amplitudes:
\bea
\label{amp31}
M_{a}^{(1)}&\approx&c\left[2m_tf\bar u\varepsilon\!\!\!/ v
+\bar u(f+if'\gamma_5)q\!\!\!/\varepsilon\!\!\!/ v\right], \qquad {\rm with}\qquad
c=\frac{g_{t\bar{t}H}g_sh}{(p_t+q)^2-M^2},\\
\label{amp32}
M_{a}^{(2)}&\approx&\bar c
\left[2m_tf\bar u\varepsilon\!\!\!/v
-\bar u\varepsilon\!\!\!/q\!\!\!/(f+if'\gamma_5) v\right], \qquad {\rm with}\qquad
\bar c=\frac{g_{t\bar{t}H}g_sh}{(p_{\bar t}+q)^2-M^2}
\eea
and for a sum of the two:
\bea
\label{amp}
M_a=M_{a}^{(1)}+M_{a}^{(2)}\approx (c+\bar c)\left[2m_tf\bar u\varepsilon\!\!\!/v
-\bar u\varepsilon\!\!\!/q\!\!\!/(f+if'\gamma_5) v \right]
+ c\; 2q\cdot\varepsilon \bar u(f+if'\gamma_5)v.
\eea
In order to calculate the sum over polarizations of the 
squared module of the matrix element  $\sum_{\rm pol.}\left|M_a\right|^2$, we take 
into account the approximate completeness relations for the spinors 
$u$ and $v$:
\bea
\sum_{\rm pol.}u \otimes \bar u \approx p_t\!\!\!\!/+m_t, \qquad
\sum_{\rm pol.}v \otimes \bar v \approx p_{\bar t}\!\!\!\!/-m_t.
\eea
and note that the off-shell polarization four vectors $\varepsilon$ are real, as 
they are defined in the following way:
\bea
\varepsilon^{\mu} \equiv 
\frac{-g^{\mu\nu}}{(p_1+p_2)^2}g_s\bar{v}(\vec{p}_1,\lambda_1)\gamma_{\nu}u(\vec{p}_2,
\lambda_2),
\eea
where the helicity spinors $v(\vec{p}_1,\lambda_1)$ and $u(\vec{p}_2,\lambda_2)$ 
of, respectively, the $\bar u$- and $u$-quark in initial state, which are calculated 
according to Eqs.~(5) and (6) of Ref.~\cite{JK}, are real if the momenta 
$\vec{p}_1$ and $\vec{p}_2$ are antiparallel. Thus
\bea
\sum_{\rm pol.}\left|M_a\right|^2&\approx&\left|(c+\bar c)\right|^2\left\{
4m_t^2f^2{\rm Tr}[(p_{\bar t}\!\!\!\!/-m_t)\varepsilon\!\!\!/(p_t\!\!\!\!/+m_t)
\varepsilon\!\!\!/]\right.\nn\\
&&\qquad\qquad  +{\rm Tr}[
(p_{\bar t}\!\!\!\!/-m_t)(f+if'\gamma_5)q\!\!\!/\varepsilon\!\!\!/(p_t\!\!\!\!/+m_t)
\varepsilon\!\!\!/q\!\!\!/(f+if'\gamma_5)]\nn\\
&&\qquad\qquad  -\left. 4m_tf{\rm Re} {\rm Tr}[
(p_{\bar t}\!\!\!\!/-m_t)\varepsilon\!\!\!/(p_t\!\!\!\!/+m_t)
\varepsilon\!\!\!/q\!\!\!/(f+if'\gamma_5)]\right\}\nn\\
&+&4|c|^2(q\cdot \varepsilon)^2{\rm Tr}[(p_{\bar t}\!\!\!\!/-m_t)(f+if'\gamma_5)
(p_t\!\!\!\!/+m_t)(f+if'\gamma_5)]\nn\\
&+&4(q\cdot \varepsilon){\rm Re}\left\{c^*(c+\bar c)\left[2m_t
f{\rm Tr}[(p_{\bar t}\!\!\!\!/-m_t)(f+if'\gamma_5)(p_t\!\!\!\!/+m_t)\varepsilon\!\!\!/]
\right.\right.\nn\\
&&\left.\left.\qquad\qquad -{\rm Tr}[(p_{\bar t}\!\!\!\!/-m_t)(f+if'\gamma_5)(p_t\!\!\!\!/+m_t)
\varepsilon\!\!\!/ q\!\!\!/(f+if'\gamma_5)]\right]\right\}.
\label{m2}
\eea
More simplified analytic form of Eq.~(\ref{m2}) is irrelevant, as the calculation
of the cross section will be performed numerically anyway, but let us note
that only the terms on the r.h.s. of Eq.~(\ref{m2}) that contain $\gamma_5$ may 
be proportional to the product $ff'$. However, if we use the relation 
$\varepsilon\!\!\!/p_t\!\!\!\!/\,\varepsilon\!\!\!/=-\varepsilon^2p_t\!\!\!\!/
+2(p_t\cdot \varepsilon)\varepsilon\!\!\!/$ in the second and third term 
and the relation 
$q\!\!\!/p_t\!\!\!\!/\,q\!\!\!/=-q^2p_t\!\!\!\!/+2(p_t\cdot q)q\!\!\!/$ in 
the second term, and then use the relation
\bea
{\rm Tr}[p_{\bar t}\!\!\!\!/\;(f+if'\gamma_5)p_t\!\!\!\!/\;
\varepsilon\!\!\!/ q\!\!\!/(f+if'\gamma_5)]=
(f^2+f'^2){\rm Tr}[p_{\bar t}\;\!\!\!\!/p_t\!\!\!\!/\;\varepsilon\!\!\!/
q\!\!\!/],
\eea
in the last term on the r.h.s. of Eq.~(\ref{m2}), 
we see that the dependence on $ff'$, and thus a sensitivity to the sign
of either $f$ or $f'$, disappears in the unpolarized cross section of the hard
scattering process 
$u \bar{u} \;\to\; b u \bar{d} \bar{b} \mu^{-} \bar{\nu}_{\mu} b \bar{b}$.
Let us note, that the same arguments can be easily repeated
for the amplitudes of the Feynman diagrams of Fig.~\ref{ggsig}, which dominate 
the $t\bar t H$ production through the gluon fusion process (\ref{ggbbbudbmn}).
We would like to stress here that all the above approximations are used
for the sake of the argument in this section only and are not used to obtain
the full results presented in Section~3.

\section{Results}

The calculation is performed in the framework of the SM, supplemented with the  
top--Higgs coupling derived from Lagrangian (\ref{Ltth}), with the use of
{\tt carlomat} \cite{kk3:C2}, a general 
purpose program for the MC computation of the lowest order cross sections.
The differential cross section of the process 
\begin{equation}
\label{pp-tt_proper}
pp \to b u \bar{d}\, \bar{b} \mu^{-} \bar{\nu}_{\mu} b \bar{b}
\end{equation}
is calculated with the use of the following factorization formula
\begin{equation}
\label{factf}
d \sigma_{pp \to b u \bar{d}\, \bar{b} \mu^{-} \bar{\nu}_{\mu} b \bar{b}}(s) 
= \sum_{i,j} \int dx_1 dx_2 \: f_i (x_1, Q^2) \: 
f_j (x_2, Q^2) \: 
d \sigma_{ij \to b u \bar{d}\, \bar{b} \mu^{-} \bar{\nu}_{\mu} b \bar{b}}(s'),
\end{equation}
\begin{figure}[htb]
\centerline{
\includegraphics[width=0.5\textwidth]{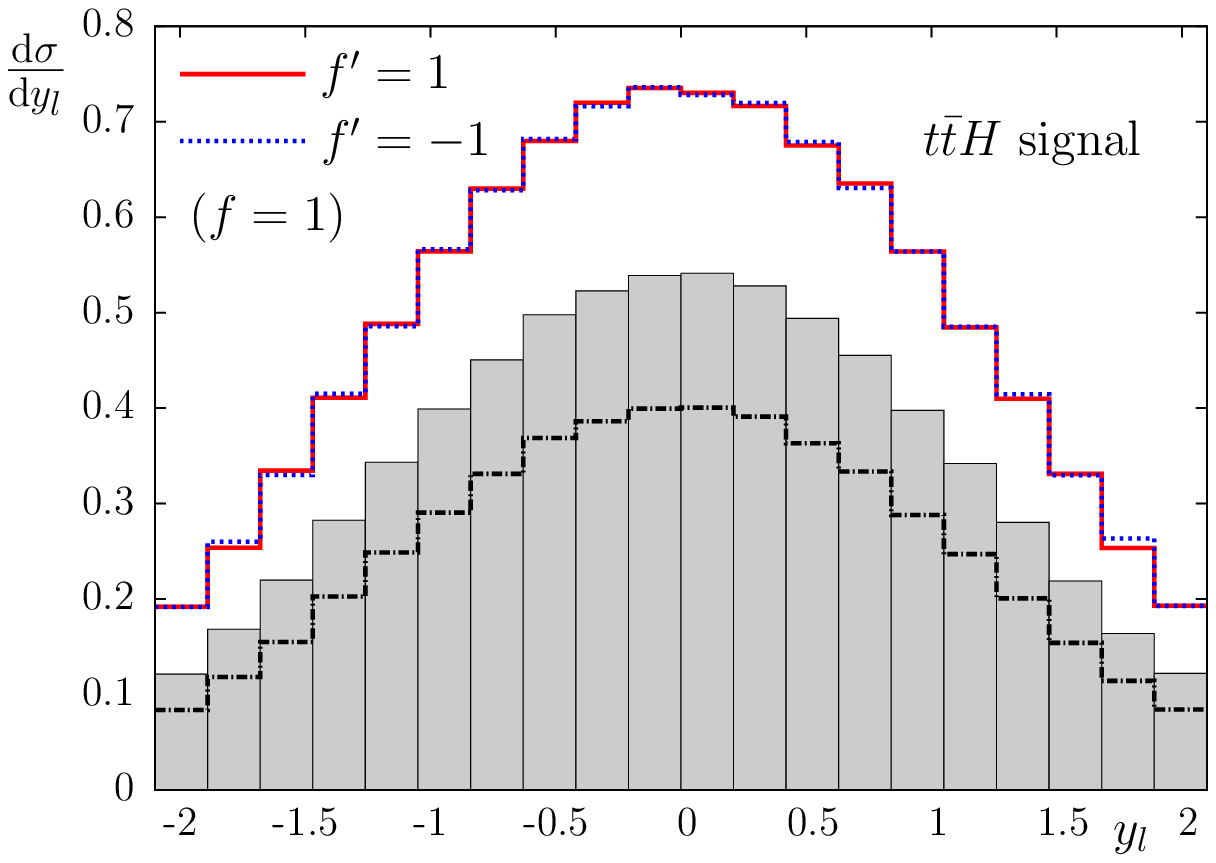}
\includegraphics[width=0.5\textwidth]{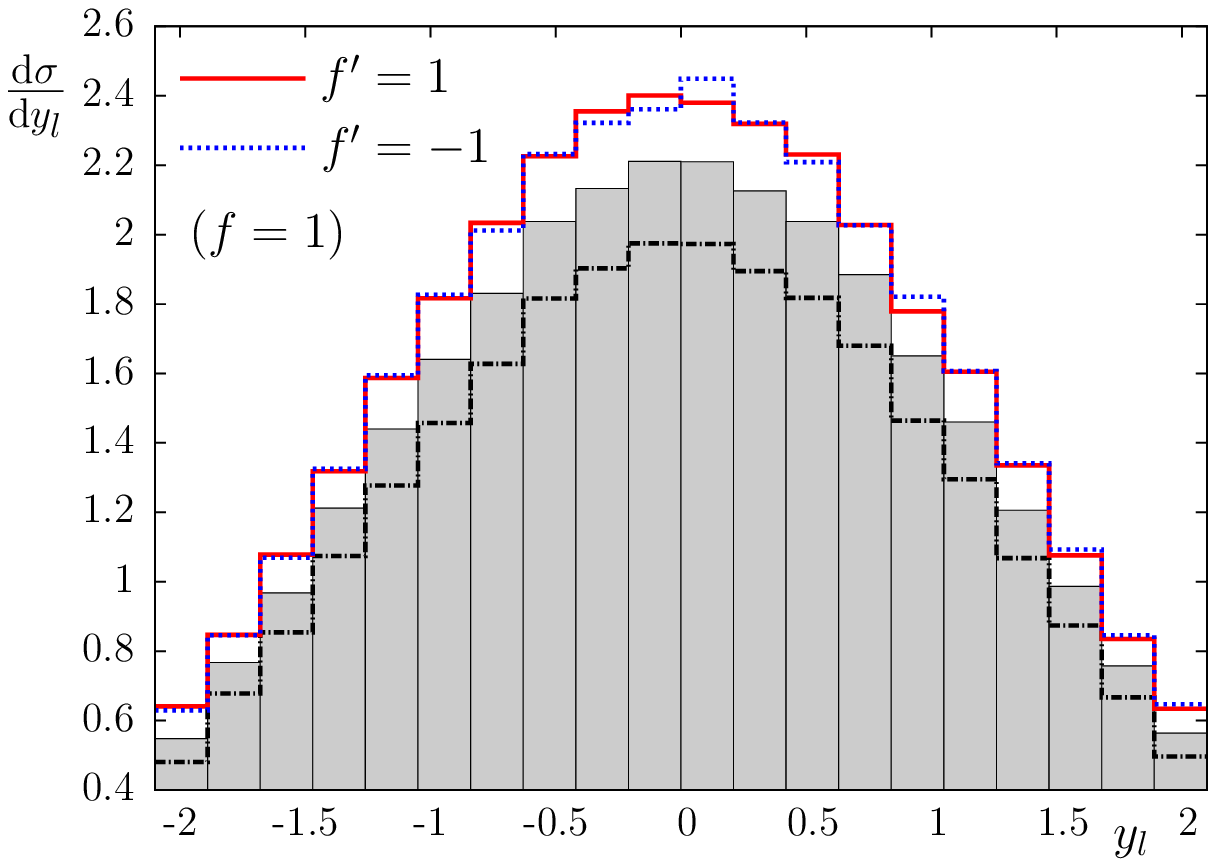}}
\caption{The differential cross section in fb of process (\ref{pp-tt_proper}) at 
$\sqrt{s}=14$~TeV as a function of the lepton rapidity computed with the 
$t\bar t H$ signal diagrams (left panel) and with all the LO diagrams (right 
panel). The SM cross section is plotted with grey shaded boxes and the 
contribution of the gluon fusion to it with the dashed-dotted line and the cross 
sections in the presence of the anomalous pseudoscalar coupling $f'=1$ ($f'=-1$) 
are plotted with the solid (dotted) line.}
\label{rapl}
\end{figure}

\begin{figure}[htb]
\centerline{
\includegraphics[width=0.5\textwidth]{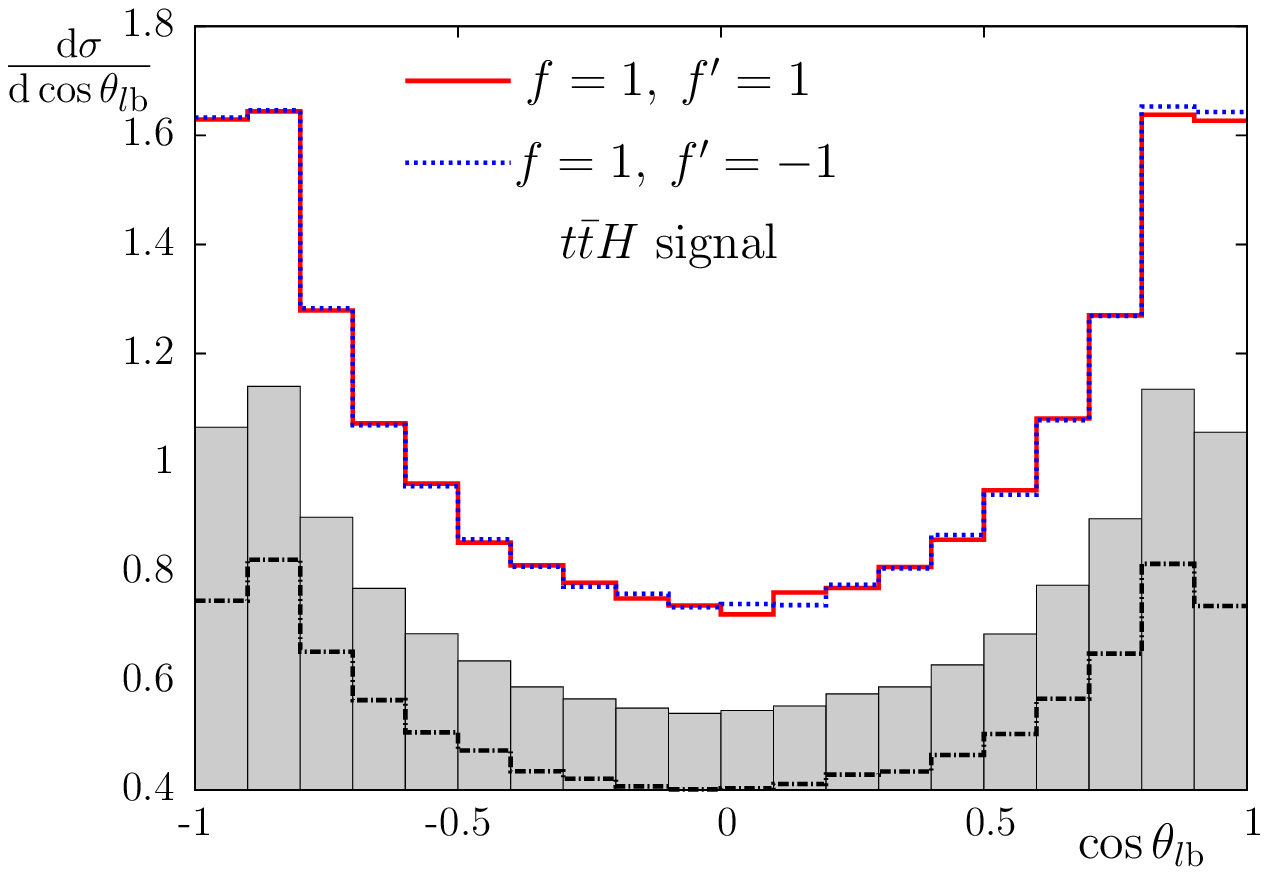}
\includegraphics[width=0.5\textwidth]{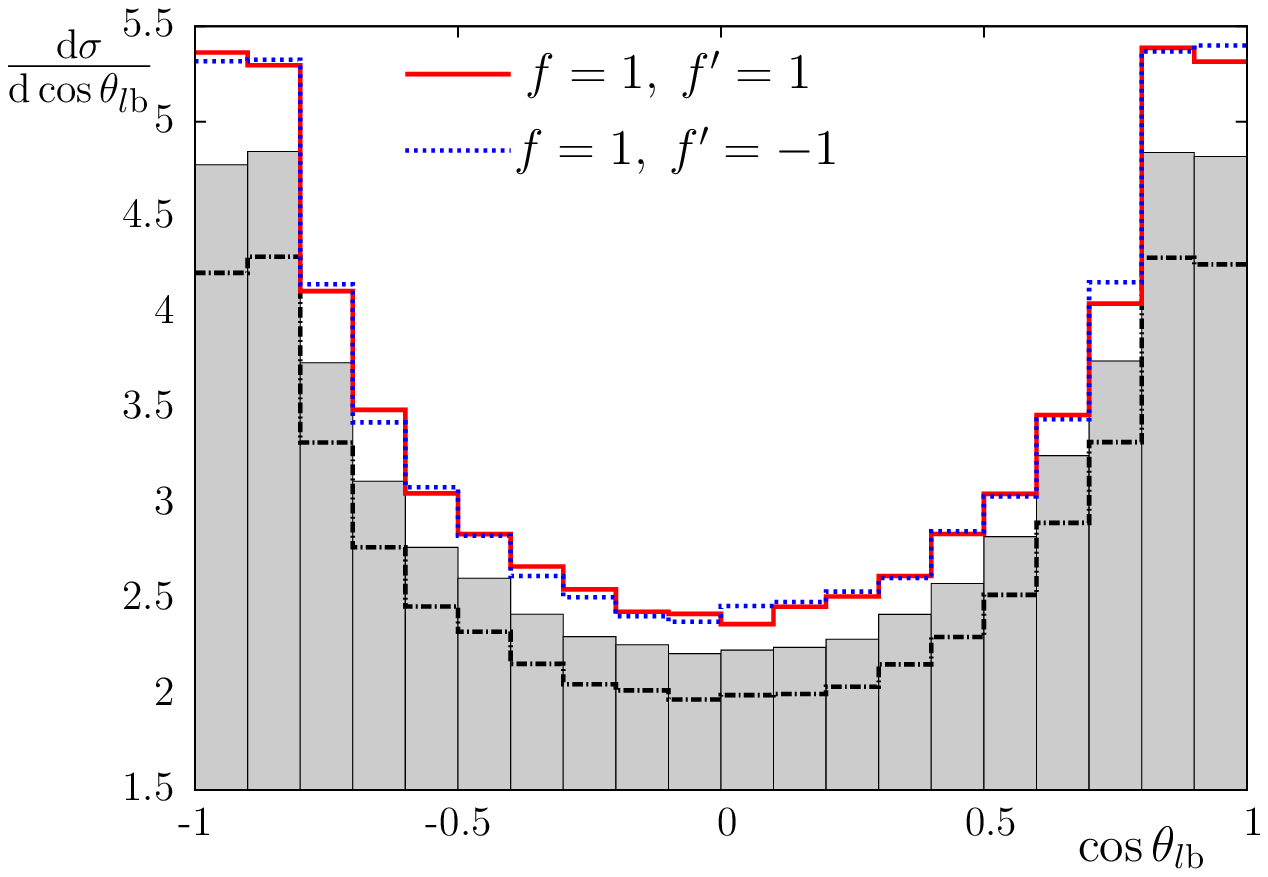}}
\caption{Same as Fig.~\ref{rapl} but as a function of the cosine of the lepton 
angle with respect to the beam.}
\label{costh_lb}
\end{figure}
where $x_1$ and $x_2$ are the proton momentum fractions carried by partons 
$i$ and $j$, respectively, $s' = x_1 x_2 s$ is the reduced center of mass energy 
squared, $Q$ is the factorization scale and we take into account the following 
pairs of partons $(i,j)$: $(g,g)$, $(u,\bar u)$, $(\bar u,u)$, $(d,\bar d)$, 
$(\bar d,d)$ in the sum. We use MSTW LO parton distribution functions \cite{MSTW} 
at the factorization scale $Q=\sqrt{m_t^2+\sum_jp_{Tj}^2}$, where $p_{Tj}$ is the 
transverse momentum of the final state quark or antiquark of process 
(\ref{pp-tt_proper}). The calculation is performed separately for the gluon 
fusion (\ref{ggbbbudbmn}) and each of the quark--antiquark hard scattering 
processes (\ref{qqbar}). We use the same physical input parameters 
and cuts (3.2)--(3.7), with $m_{bb}^{\rm cut}=20$~GeV in (3.7), as in 
Ref.~\cite{kkjhep}, and three different combinations of the scalar and 
pseudoscalar couplings of Lagrangian (\ref{Ltth}): $(f,f')=(1,0),(1,1),(1,-1)$. 
The first combination corresponds to the SM and the other two are chosen, just 
for the sake of illustration, beyond the allowed 95\%~CL regions of the $(f,f')$ 
plane which, as discussed in the first paragraph of Section~2, are model 
dependent anyway. The cross sections of the hard scattering processes considered 
are added afterwards, if necessary.

\begin{figure}[htb]
\centerline{
\includegraphics[width=0.5\textwidth]{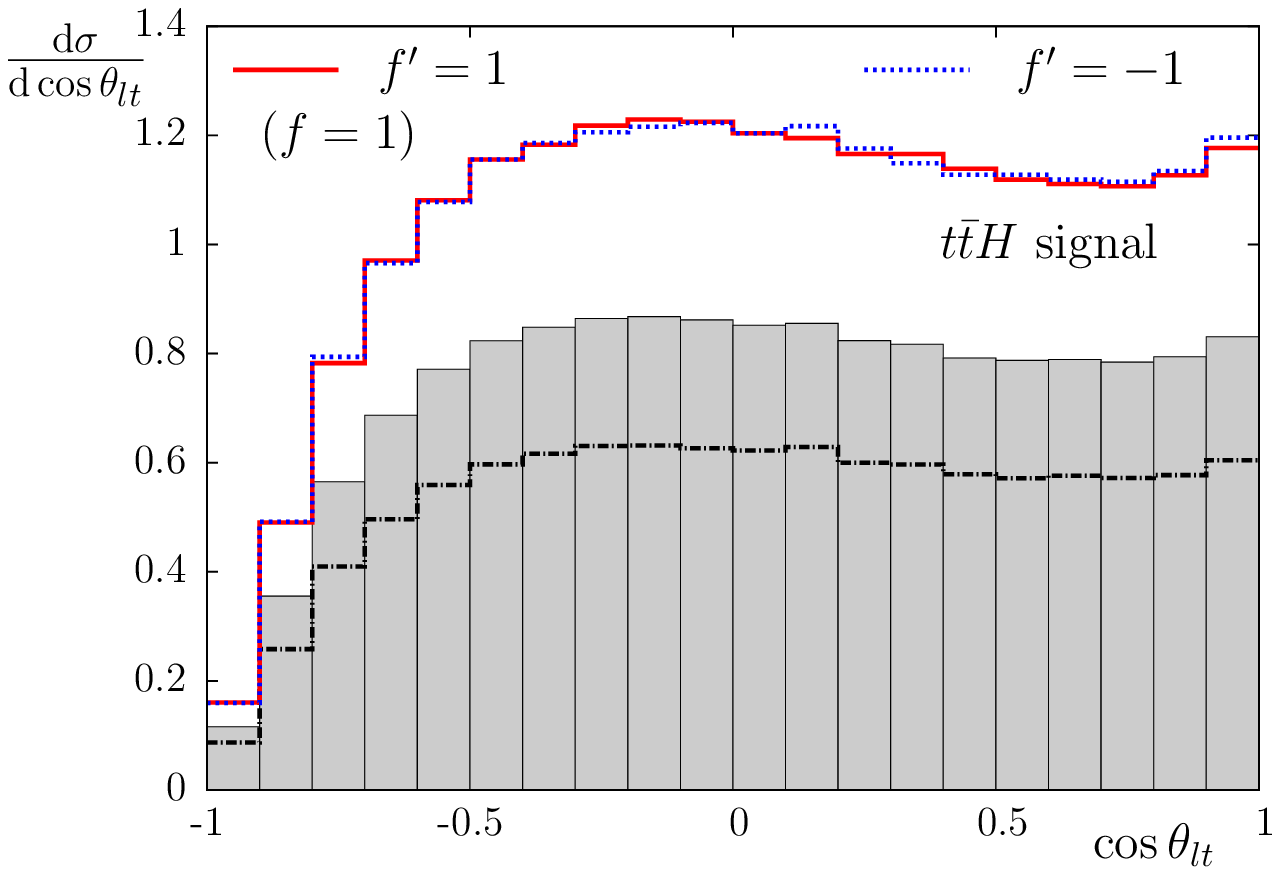}
\includegraphics[width=0.5\textwidth]{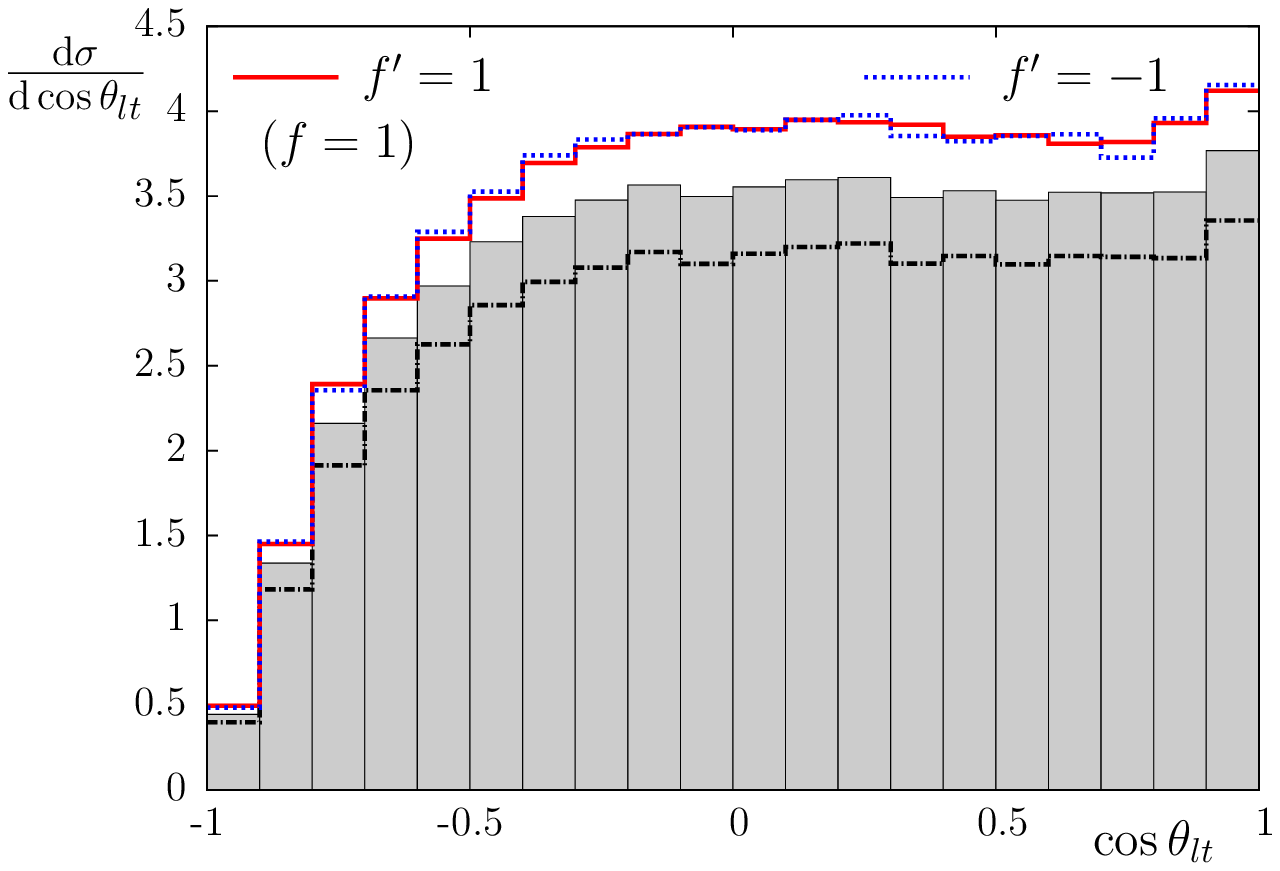}}
\caption{Same as Fig.~\ref{rapl} but as a function of the cosine of the lepton 
angle with respect to the top quark in the top quark rest frame.}
\label{costh_lt}
\end{figure}

\begin{figure}[htb]
\centerline{
\includegraphics[width=0.5\textwidth]{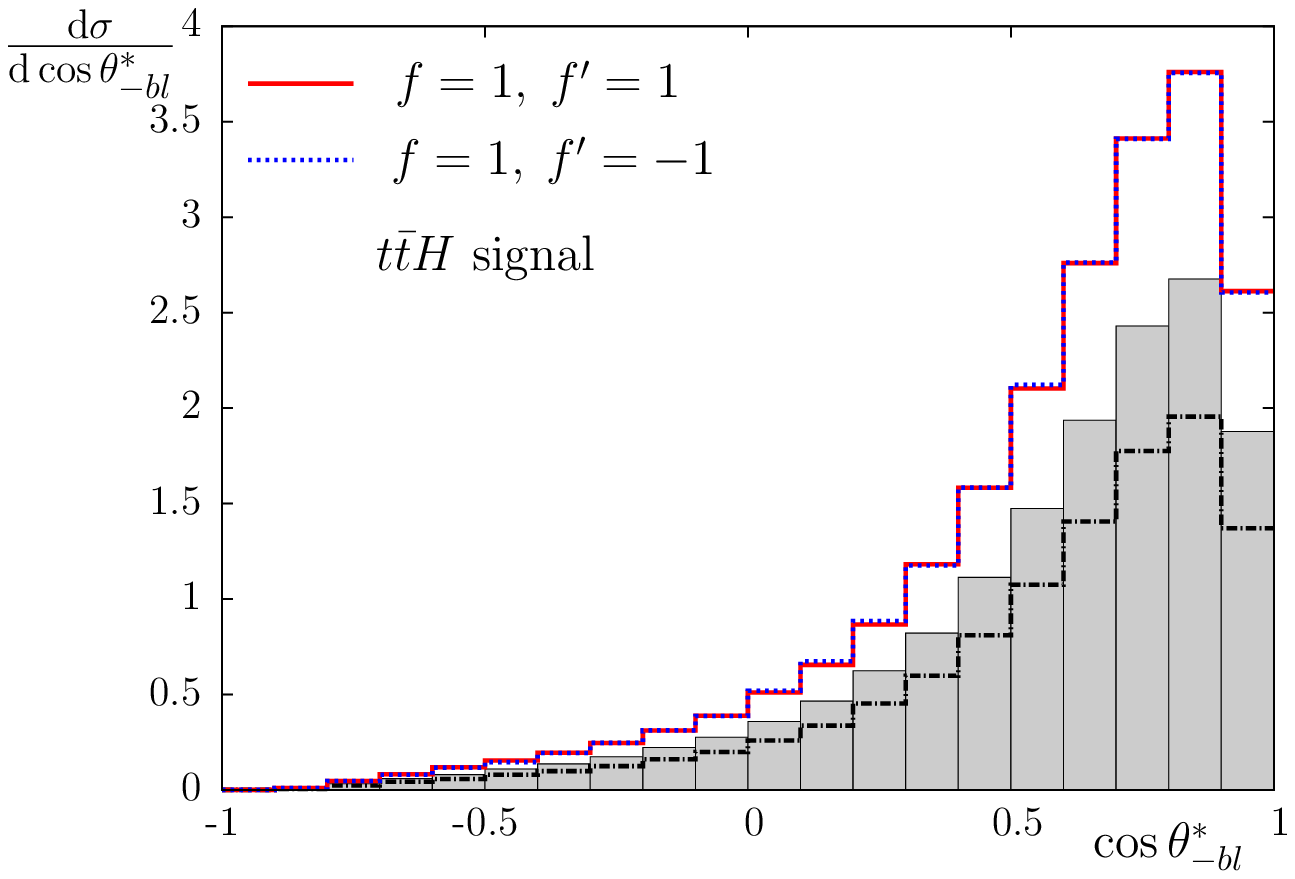}
\includegraphics[width=0.5\textwidth]{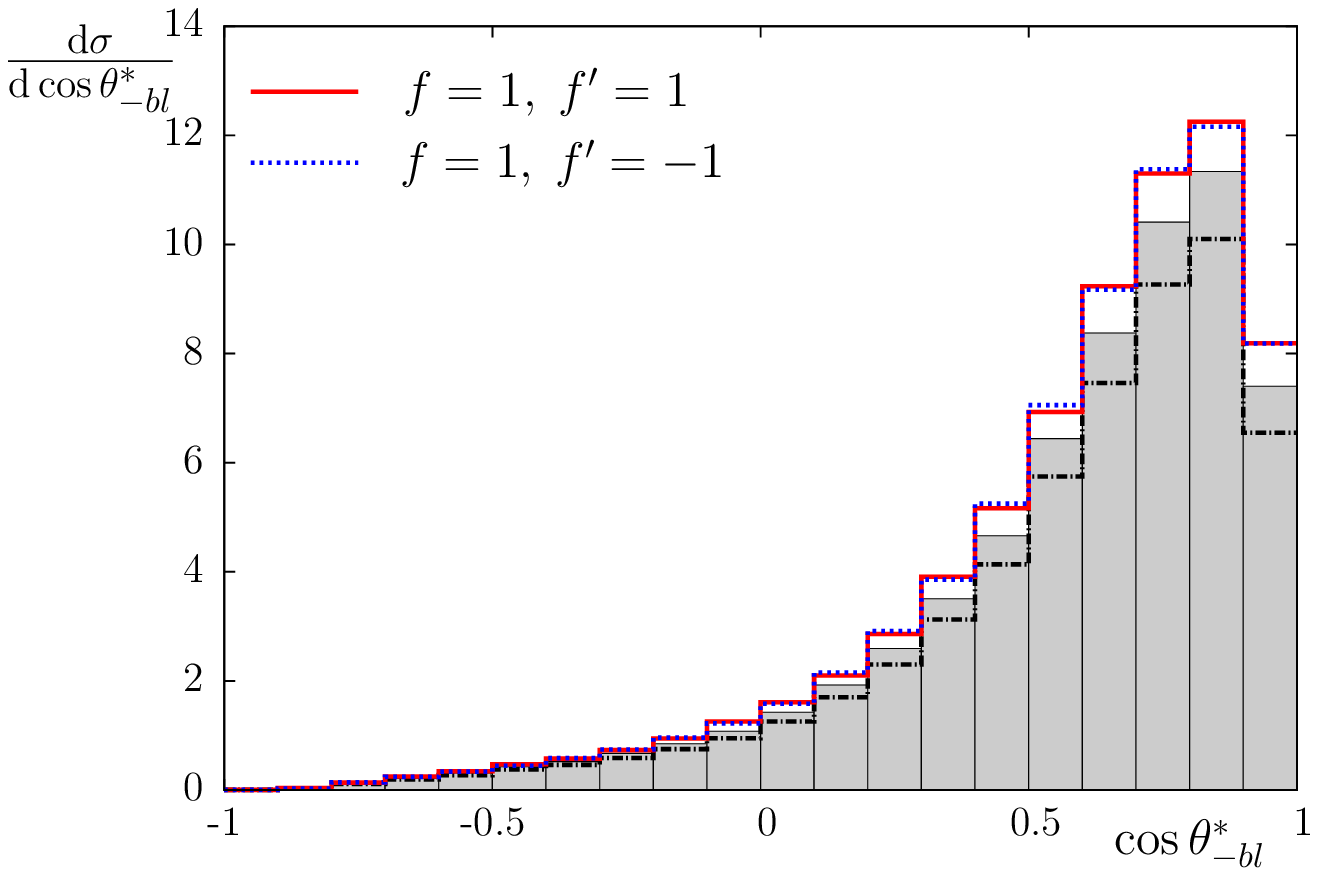}}
\caption{Same as Fig.~\ref{rapl} but as a function of the cosine of the lepton 
angle with respect to the inverse momentum of the $b$-quark in the top quark 
rest frame.}
\label{costh_lmb}
\end{figure}

Let us note, that in order to calculate the total cross section of process 
(\ref{pp-tt_proper}), a 20-fold phase space integral and a 2-fold integral over 
parton density functions must be performed, not to mention the additional 9-fold 
Monte Carlo (MC) integral that replaces the sum over particle helicities,
without which the computation would not have been feasible in practice. 

\begin{figure}[htb]
\centerline{
\includegraphics[width=0.5\textwidth]{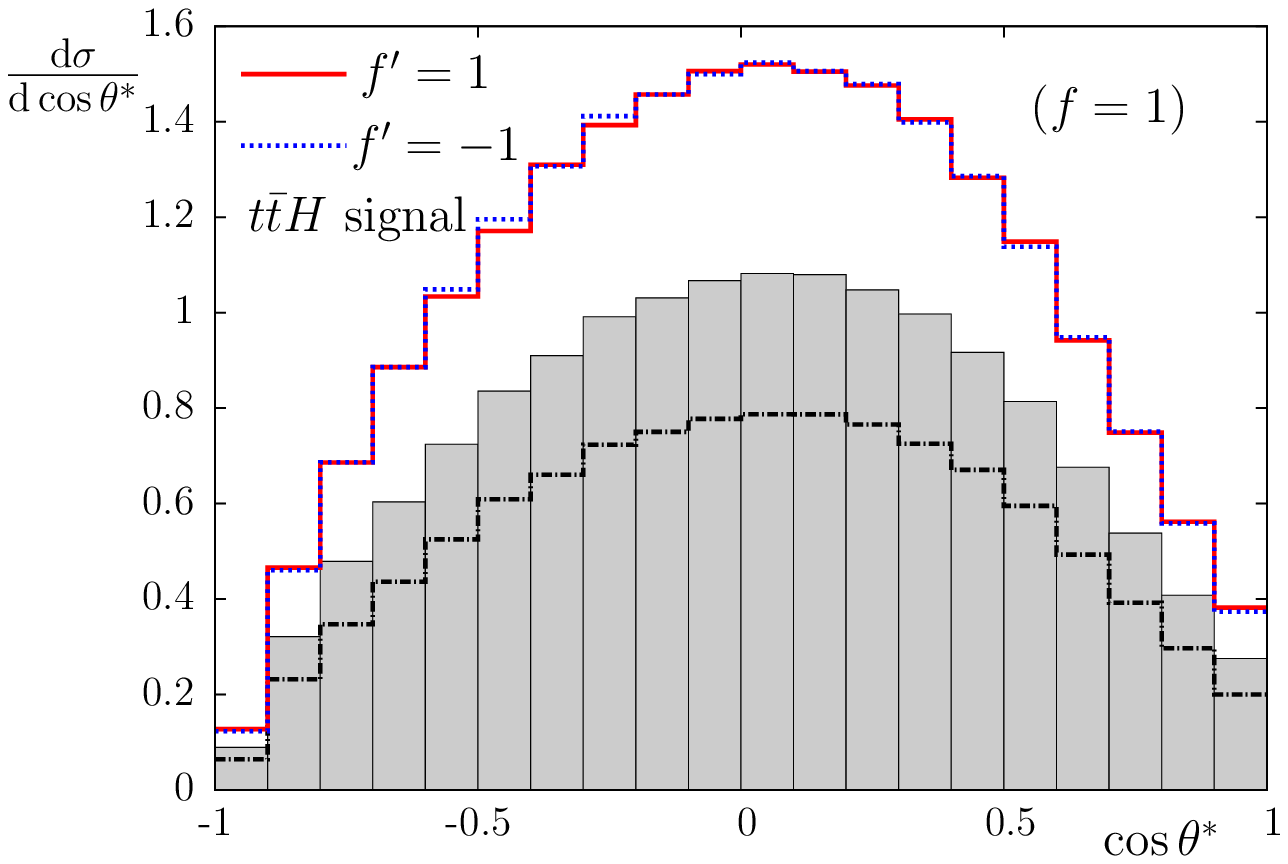}
\includegraphics[width=0.5\textwidth]{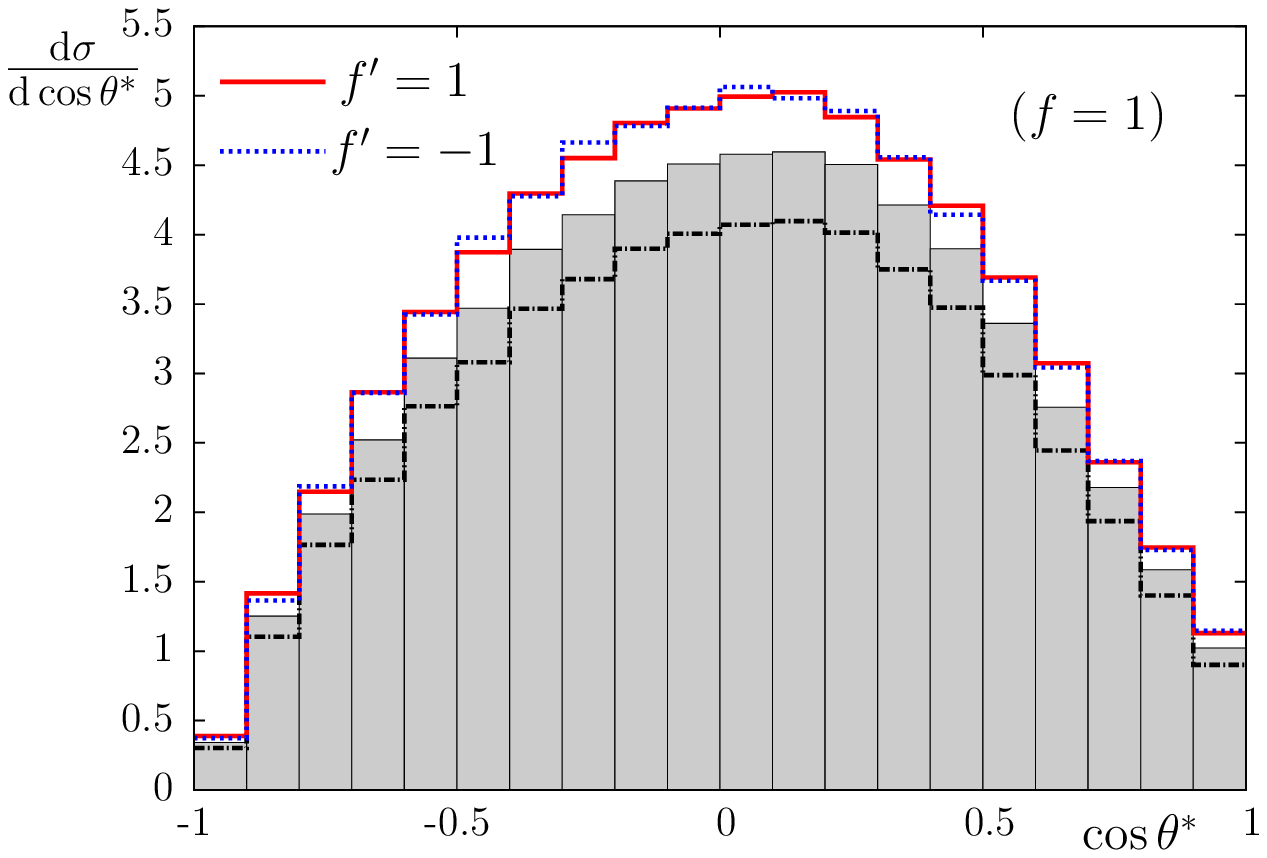}}
\caption{Same as Fig.~\ref{rapl} but as a function of the cosine of the lepton 
angle with respect to the inverse momentum of the $b$-quark in the $W^-$ boson 
rest frame.}
\label{costh_lmb_wrf}
\end{figure}

\begin{figure}[htb]
\centerline{
\includegraphics[width=0.5\textwidth]{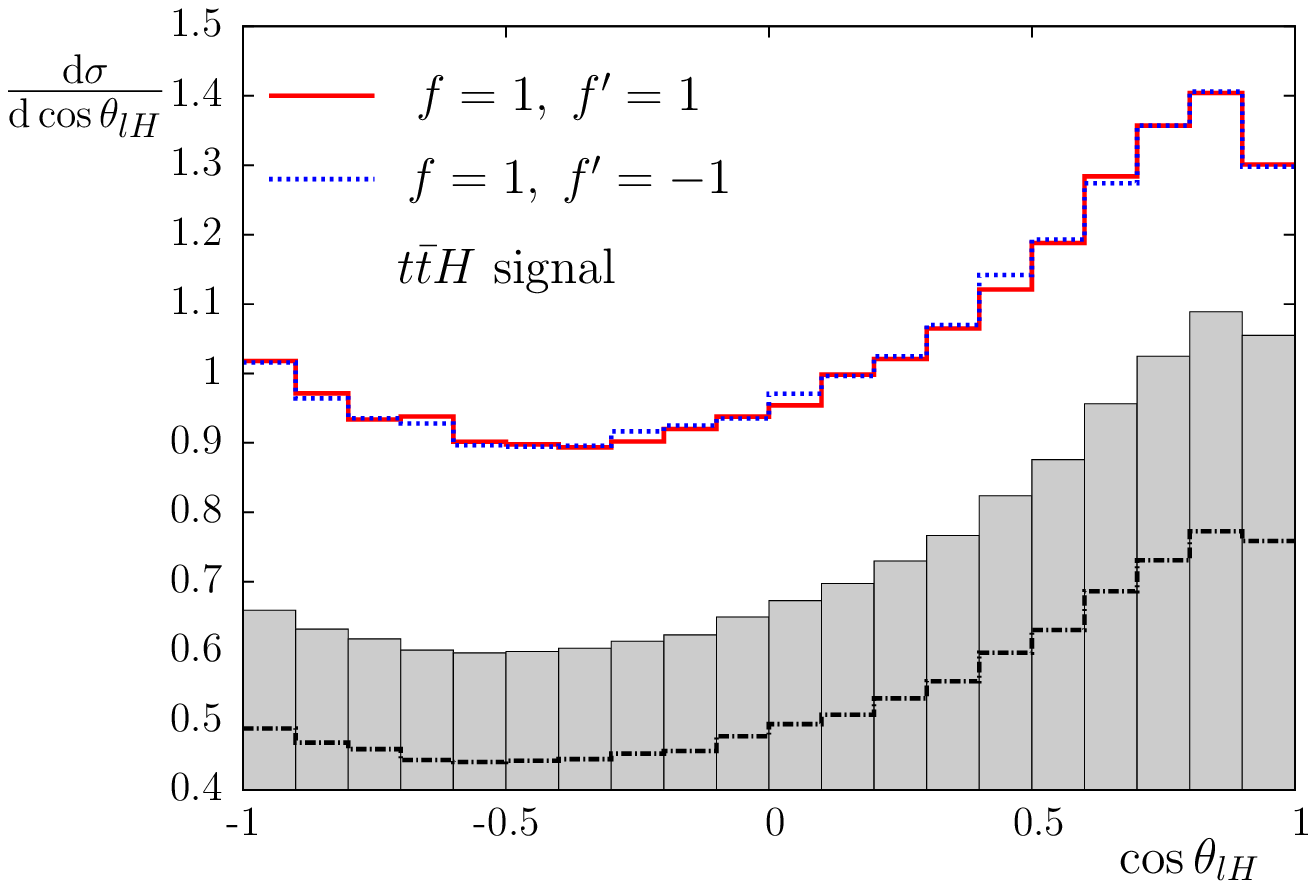}
\includegraphics[width=0.5\textwidth]{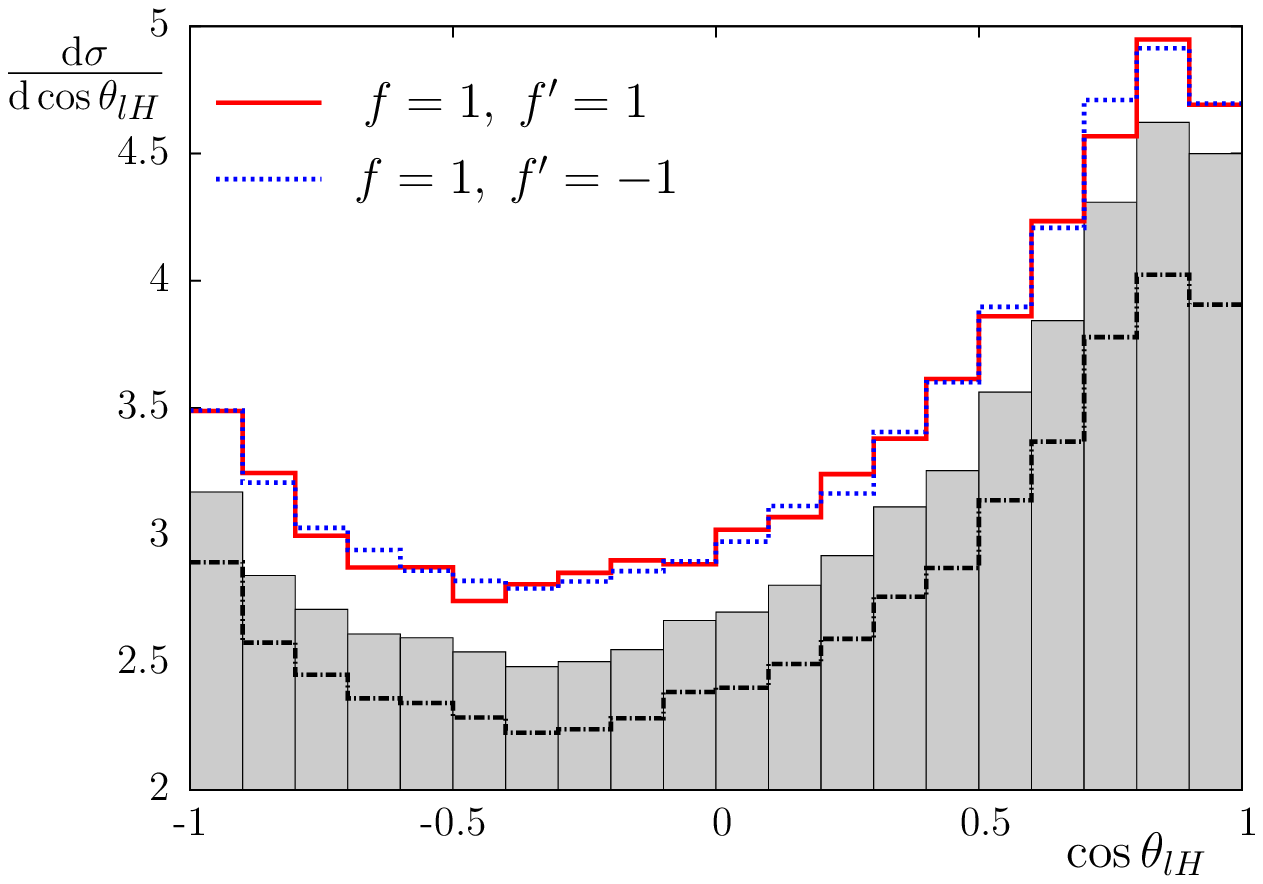}}
\caption{Same as Fig.~\ref{rapl} but as a function of the cosine of the lepton 
angle with respect to the Higgs boson in the LAB frame.}
\label{costh_hl}
\end{figure}

\begin{figure}[htb]
\centerline{
\includegraphics[width=0.5\textwidth]{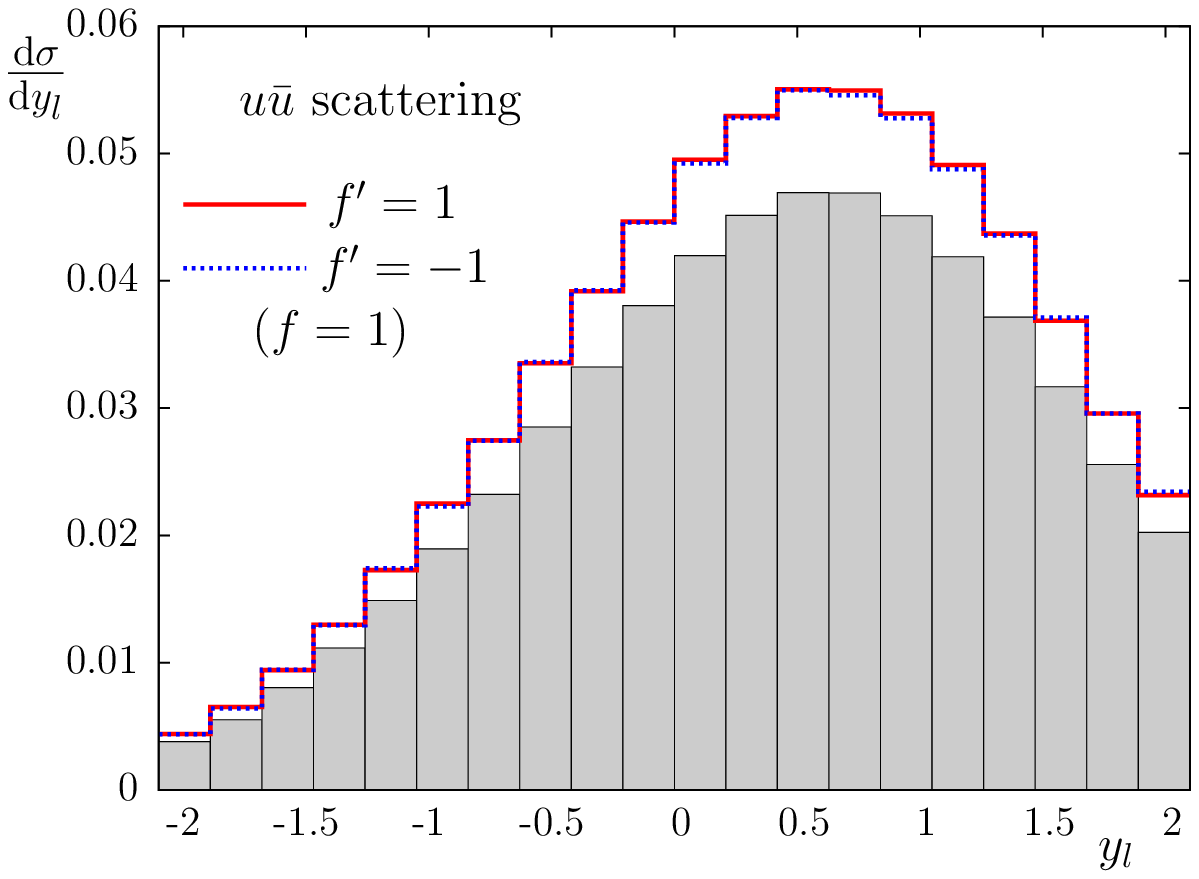}
\includegraphics[width=0.5\textwidth]{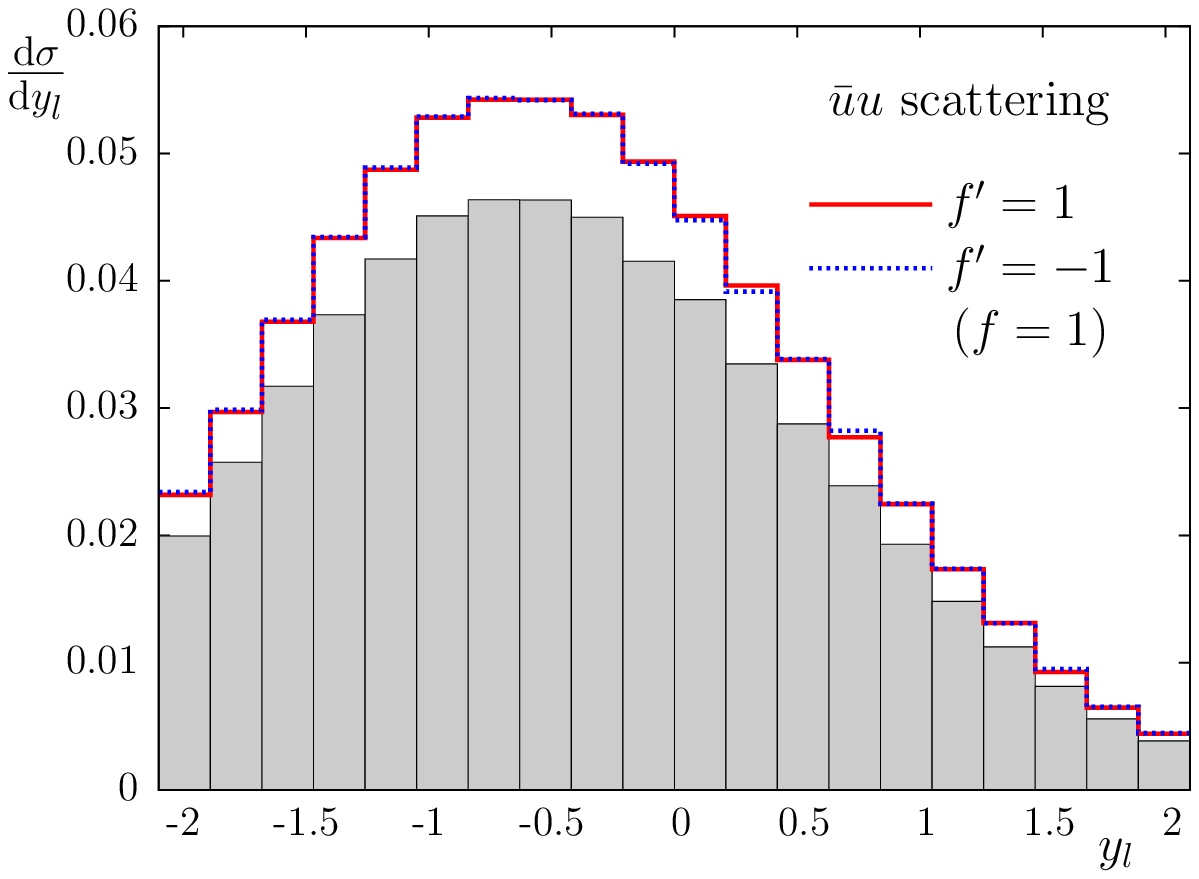}}
\caption{The differential cross section in fb of process (\ref{pp-tt_proper}) at 
$\sqrt{s}=14$~TeV as a function of the lepton rapidity computed with the 
$t\bar t H$ signal diagrams of the $u\bar u$- (left panel)
and $\bar u u$-hard-scattering processes (right panel). The corresponding
SM cross section is plotted with grey shaded boxes.}
\label{rapl_uu}
\end{figure}

\begin{figure}[htb]
\centerline{
\includegraphics[width=0.5\textwidth]{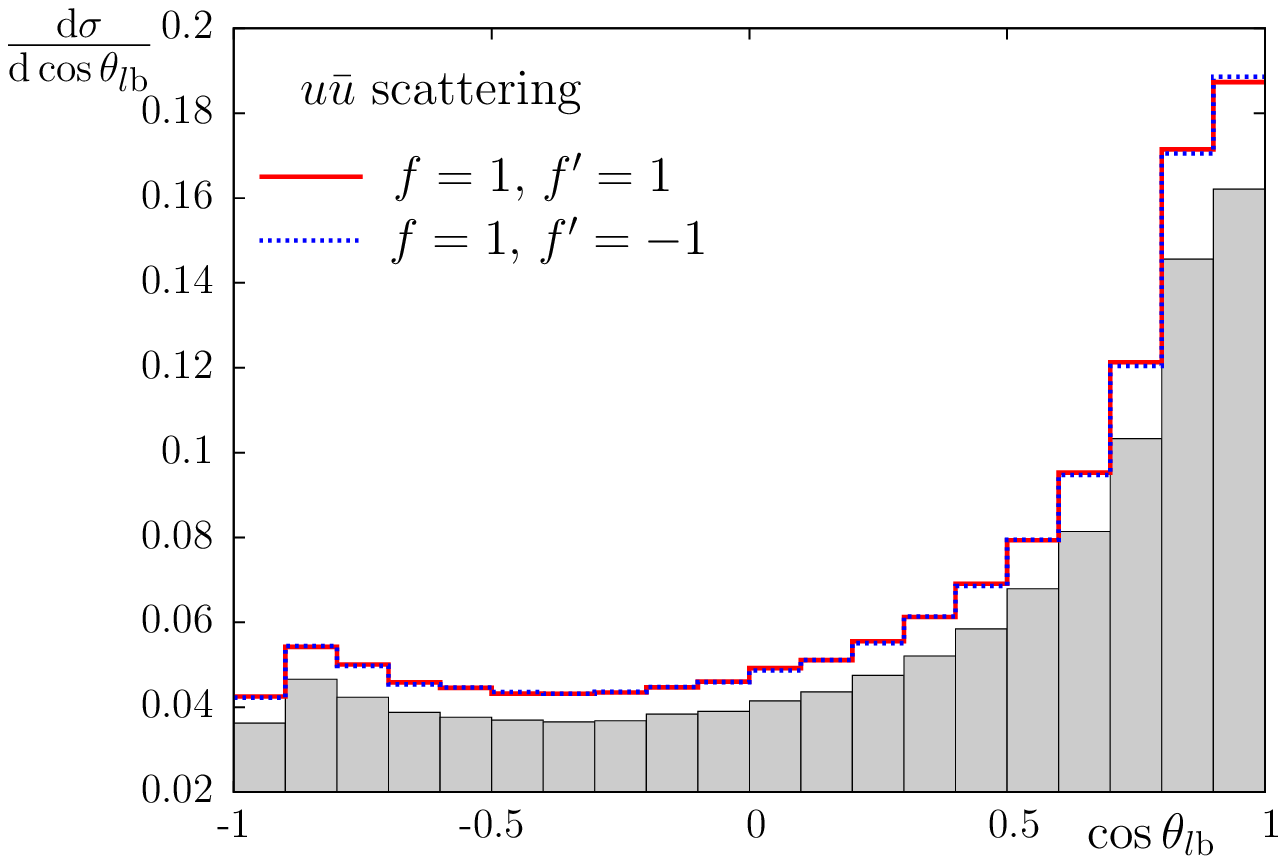}
\includegraphics[width=0.5\textwidth]{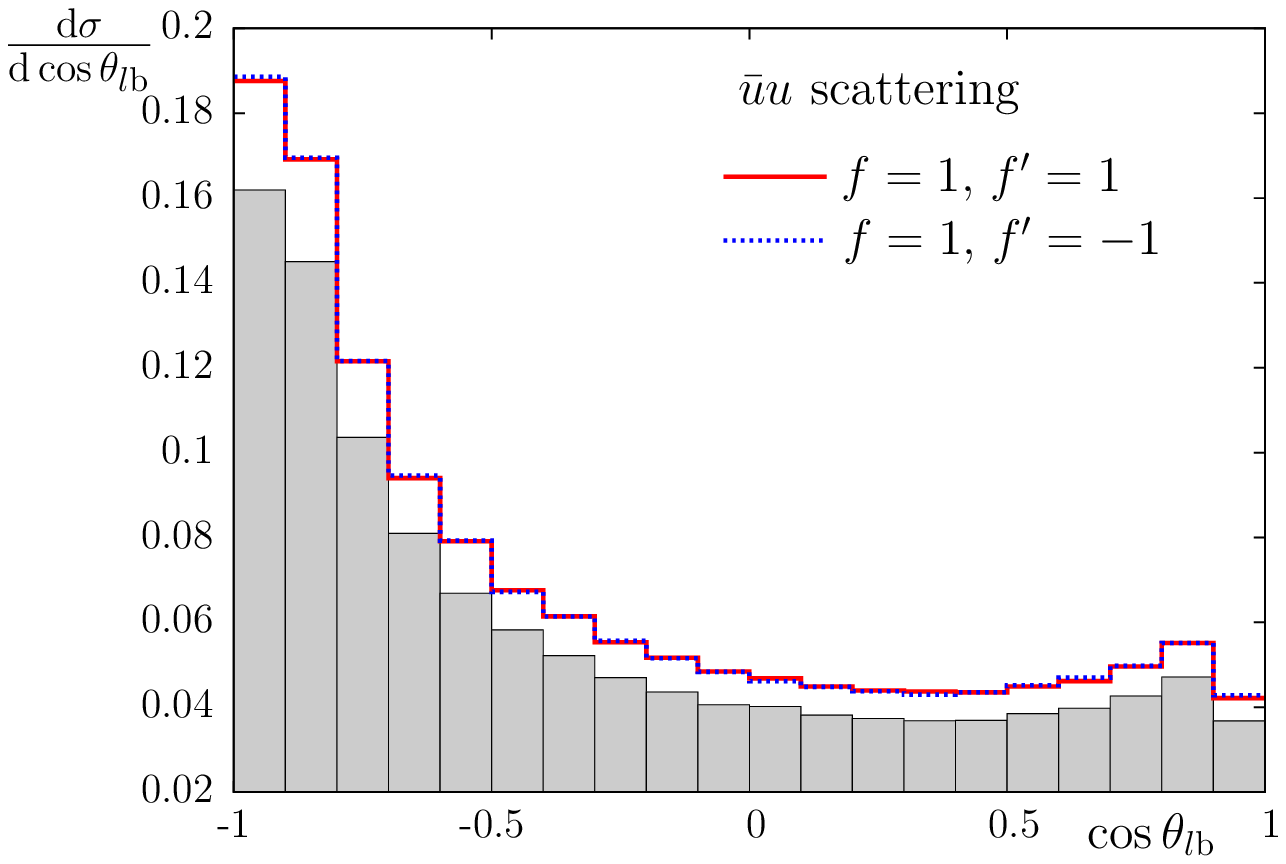}}
\caption{Same as Fig.~\ref{rapl_uu} but as a function of the cosine of the 
lepton angle with respect to the beam.}
\label{costh_lb_uu}
\end{figure}

\begin{figure}[htb]
\centerline{
\includegraphics[width=0.5\textwidth]{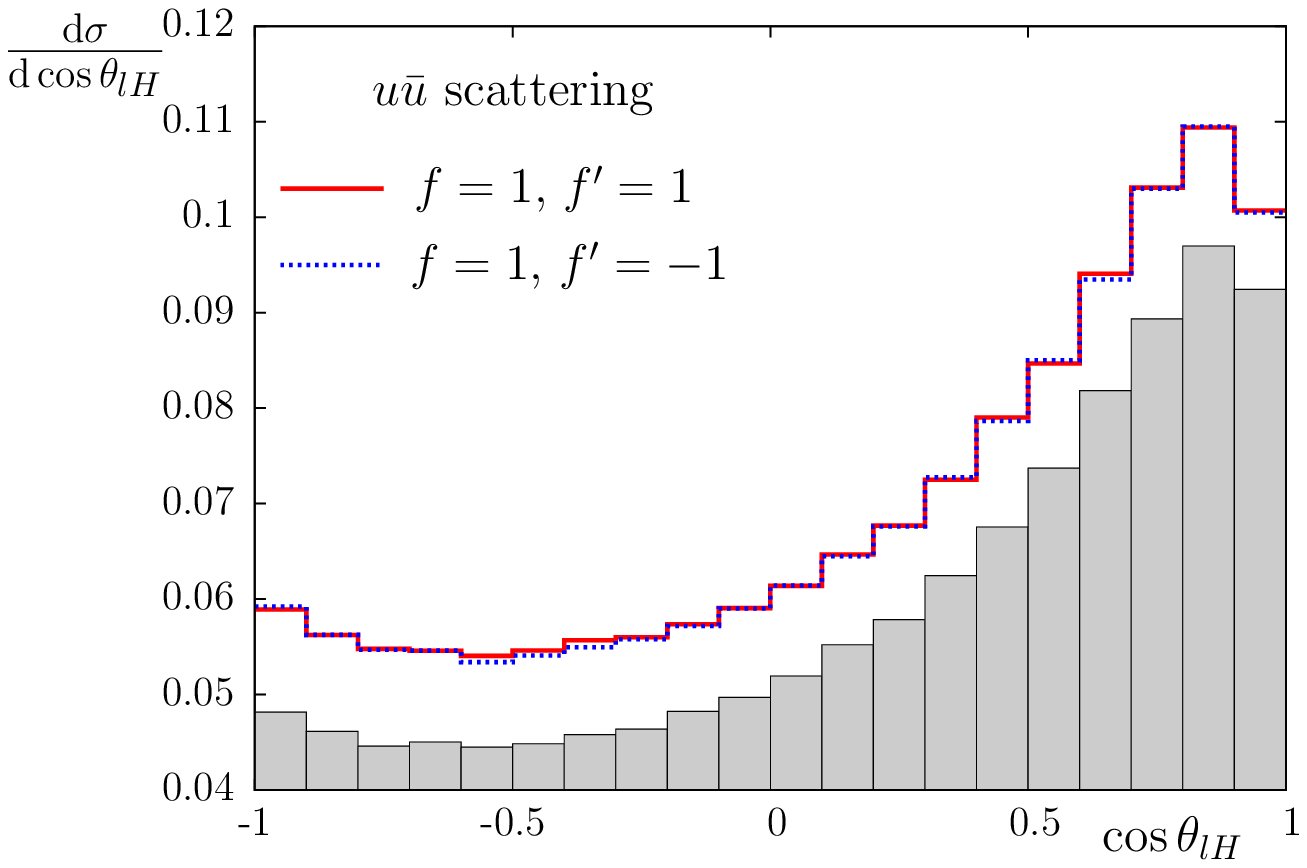}
\includegraphics[width=0.5\textwidth]{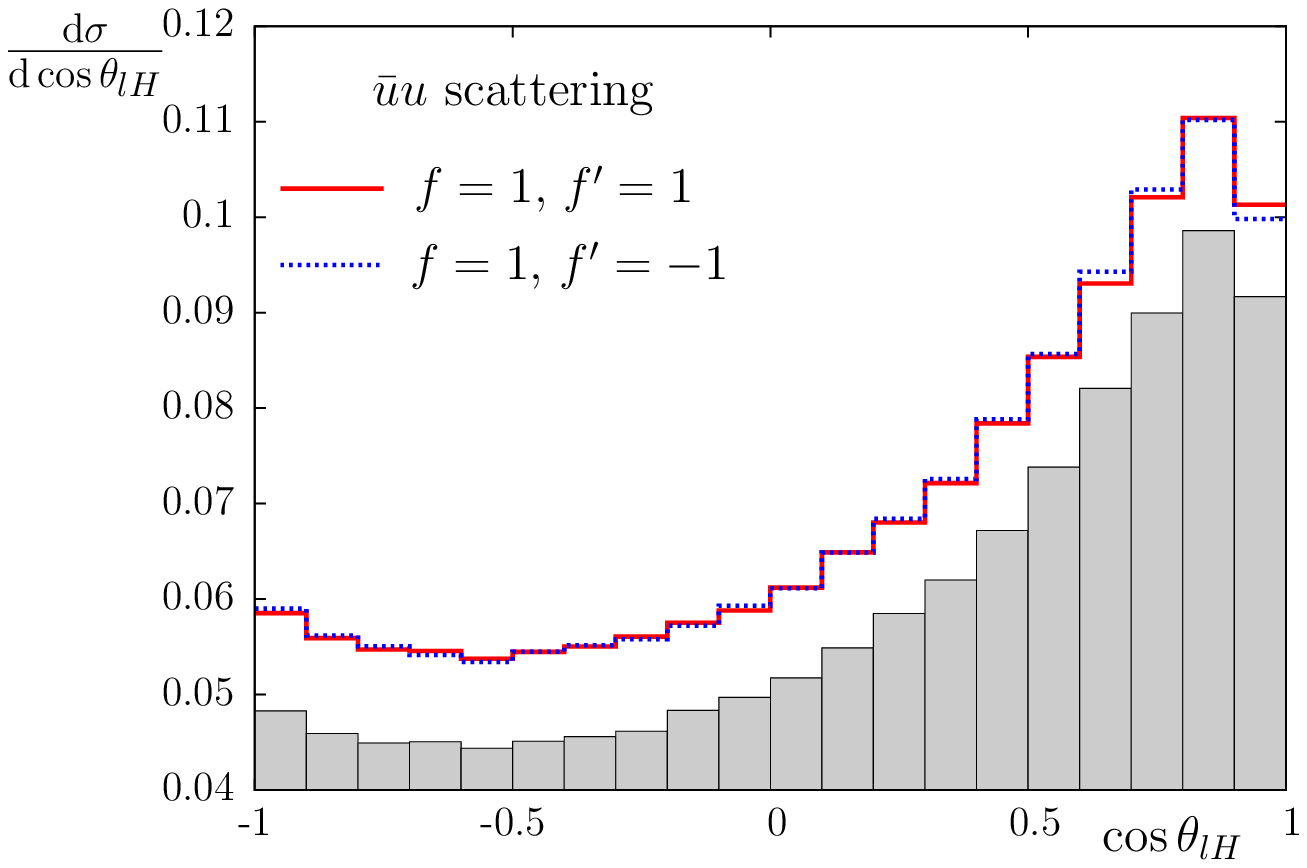}}
\caption{Same as Fig.~\ref{rapl_uu} but as a function of the cosine of the 
lepton angle with respect to the Higgs boson in the LAB frame.}
\label{costh_hl_uu}
\end{figure}

The differential cross sections of process (\ref{pp-tt_proper}) at the 
proton--proton center of mass energy of 14~TeV are plotted in 
Figs.~\ref{rapl}--\ref{costh_hl} as functions of the rapidity and different 
angular variables of the final state muon, being referred to as the lepton.
In Figs.~\ref{rapl}--\ref{costh_hl}, the left panels show the signal cross 
sections, which are computed with the signal $t\bar tH$ production diagrams 
of the hard scattering processes (\ref{ggbbbudbmn}) and 
(\ref{qqbar}), as described in the previous sections,
and the right panels show the complete LO cross sections, which are computed 
with the complete set of the LO Feynman diagrams of each of the hard scattering 
processes considered.
In each of the figures, the SM cross section of process (\ref{pp-tt_proper}) is 
plotted with grey shaded boxes and the contribution of the gluon fusion to it 
with the dashed-dotted line and the cross sections in the presence of the 
anomalous pseudoscalar coupling $f'=1$ ($f'=-1$) are plotted with the solid 
(dotted) line. Thus, the shaded area above the dashed-dotted line
shows the contribution of the quark--antiquark hard scattering processes to 
either the $t\bar tH$ signal or complete SM cross section.
The effects of the anomalous pseudoscalar coupling $f'=\pm 1$ are quite 
sizable in the signal cross sections which become by about 50\% bigger than 
in the SM. If all the LO Feynman diagrams are taken into account the effects 
remain the same in absolute terms, but their relative size is substantially 
smaller, as the anomalous top--Higgs coupling (\ref{Ltth}) practically does not 
alter the off resonance background contributions 
which substantially increase the cross section of process (\ref{pp-tt_proper}).
The shape of each of the differential cross sections plotted in 
Figs.~\ref{rapl}--\ref{costh_hl} is hardly changed in the presence of the 
anomalous coupling $f'=\pm 1$. Moreover, the cross sections for $f'=1$
and $f'=-1$ look almost identical, which means that the process is practically 
insensitive to a sign of $f'$, in accordance with the discussion of Section~2.

The individual contributions of the $u\bar u$- and $\bar u u$-hard-scattering 
processes to the $t\bar t H$ signal differential cross sections of process 
(\ref{pp-tt_proper}) at $\sqrt{s}=14$~TeV are plotted in Figs.~\ref{rapl_uu}, 
\ref{costh_lb_uu} and \ref{costh_hl_uu}, as functions of the lepton 
rapidity, cosine of the lepton angle with respect to the beam and cosine of the 
lepton angle with respect to the Higgs boson in the laboratory (LAB) frame, 
respectively.
The relative effects of the anomalous pseudoscalar coupling $f'$ in the plots 
of Figs.~\ref{rapl_uu}, \ref{costh_lb_uu} and \ref{costh_hl_uu} are approximately 
the same as in the full signal cross sections plotted in the right panels of 
Figs.~\ref{rapl}, \ref{costh_lb} and \ref{costh_hl}, respectively, and again 
there is practically no sensitivity to the sign of $f'$. 
Taking into account the off resonance background contributions
to any of the quark--antiquark hard scattering processes does not change this
conclusion either, i. e., the shapes and relative effect of the anomalous 
coupling $f'$ remain practically the same for all the distributions considered.  

\section{Conclusions}

We have complemented the analysis of the influence of the 
anomalous Higgs boson coupling to top quark on the secondary lepton 
distributions in the process of associated production of the top quark pair 
and Higgs boson in the proton--proton collisions at the LHC of Ref.~\cite{kkjhep}
by taking into account contributions of the quark--antiquark annihilation hard 
scattering processes.
Although, the gluon fusion mechanism dominates the $t\bar t H$ production 
through process (\ref{pp-tt_proper}) at $\sqrt{s}=14$~TeV, the 
contribution of quark--antiquark hard scattering processes (\ref{qqbar})
is quite substantial and, therefore, should be taken into account in the analyses 
of data. Moreover, we have explained why the effects of the scalar and pseudoscalar
anomalous couplings in the unpolarized cross section of the process are completely 
insensitive to the sign of either of them.

\FloatBarrier

\end{document}